\documentclass[12pt]{article}
\usepackage{hyperref}
\usepackage{cite}
\usepackage{xcolor}
\usepackage{graphicx}
\usepackage{caption,subcaption}
\usepackage{amsmath}
\usepackage{amssymb}
\usepackage{xspace}
\usepackage{verbatim}
\usepackage{mathtools}
\usepackage[normalem]{ulem}

\makeatletter
\@addtoreset{equation}{section}

\makeatletter
\renewcommand\section{\@startsection {section}{1}{\z@}%
                                   {-3.5ex \@plus -1ex \@minus -.2ex}
                                   {2.3ex \@plus.2ex}%
                                   {\normalfont\large\bfseries}}
\renewcommand\subsection{\@startsection{subsection}{2}{\z@}%
                                     {-3.25ex\@plus -1ex \@minus -.2ex}%
                                     {1.5ex \@plus .2ex}%
                                     {\normalfont\bfseries}}

\def\baselinestretch{1.2}
\newcommand{\be}{\begin{equation}}
\newcommand{\ee}{\end{equation}}
\newcommand{\beq}{\begin{eqnarray}}
\newcommand{\eeq}{\end{eqnarray}}

\newcommand{\gone}[1]{{}}

\begin{document}
\begin{titlepage}
\begin{flushright}
MAD-TH-16-04
\end{flushright}

\vfil

\begin{center}

{\bf \Large
Tunneling in Axion Monodromy
}

\vfil

Jon Brown, William Cottrell, Gary Shiu, and Pablo Soler

\vfil

Department of Physics, University of Wisconsin, Madison, WI 53706, USA

\vfil
\end{center}

\begin{abstract}

\noindent The Coleman formula for vacuum decay and bubble nucleation has been used to estimate the tunneling rate in models of axion monodromy in recent literature.  However, several of Coleman's original assumptions do not hold for such models.  Here we derive a new estimate with this in mind using a similar Euclidean procedure.  We find that there are significant regions of parameter space for which the tunneling rate in axion monodromy is not well approximated by the Coleman formula.  However, there is also a regime relevant to large field inflation in which both estimates parametrically agree.  We also briefly comment on the applications of our results to the relaxion scenario.

\end{abstract}
\vspace{0.5in}

\end{titlepage}
\renewcommand{\baselinestretch}{1.05}  

\section{Introduction}

While the theory of inflation stands as an extremely successful phenomenological model, finding a UV complete description of it remains as one of the most severe challenges to any theory of quantum gravity.  Within the context of string theory, much attention has been focused on large field inflation, partially motivated by the fact that ongoing experiments will provide ever sharper constraints on this scenario.  It would be very satisfying if (string) theory could make a categorical prediction regarding large field ranges, especially {\it{before}} the issue is decided experimentally.

While short of being definitive, string theory certainly seems to disfavor explicit, well-controlled models that have large field ranges. We will focus our discussion on axions, whose discrete shift symmetry makes them a promising candidate for maintaining perturbative control \cite{Freese:1990rb}.  For this reason, much effort has been put into understanding the phenomenology of axionic or `natural' inflation.

However, for a single axion field,  tension between the requirements of small instanton corrections and large field ranges emerges when one tries to embed this scenario into string theory \cite{Banks:2003sx}.  This observation was later subsumed by the so-called `Weak Gravity Conjecture' (WGC) \cite{ArkaniHamed:2006dz}.  In its simplest form, the WGC states that the axion field range, $2\pi f$, and the instanton action, $S_\text{inst}$, are constrained by $f\times S_\text{inst} \lesssim M_{P}$.  This means that trans-Planckian field ranges are always accompanied by unsuppressed instantons.

It was furthermore established in~\cite{Rudelius:2015xta,Brown:2015iha,Brown:2015lia} that scenarios involving multiple axions such as N-flation~\cite{Dimopoulos:2005ac} and alignment~\cite{Kim:2004rp}  would fail to suppress higher instanton corrections assuming a suitably generalized form of the WGC.  These results create substantial obstacles for many models of axionic inflation and may explain why the thorough investigations of~\cite{Bachlechner:2014gfa,Long:2016jvd} have not succeeded in finding a compelling large field model.

One possible way out is provided by axion monodromy \cite{Silverstein:2008sg,McAllister:2008hb,Kaloper:2008fb,Kaloper:2011jz,Marchesano:2014mla}, which maintains the axion's perturbative control while circumventing restrictions on the field range.  More precisely, in axion monodromy a perturbative potential is induced on top of the usual cosine potential.  This allows the axion to traverse through multiple field ranges during inflation.  Thus, $f$ may be kept sub-Planckian and the WGC does not prevent the instantonic corrections from being sufficiently small.  For these reasons, axion monodromy has become one of the dominant inflationary paradigms within the recent string theory literature.

Axion monodromy, however, is susceptible to a new kind of problem: membrane nucleation \cite{Kaloper:2011jz}.  More precisely, it is possible for a membrane to nucleate during inflation and mediate a spontaneous decay to a lower energy state.  This is possible due to the multi-branch structure of the potential
\begin{equation}
	V\left(\phi\right) = \frac{1}{2}\mu^2\left(\phi + \frac{ne}{\mu}\right)^2+\Lambda_\text{inst}^{4} e^{-S_{\text{inst}}} \left(1-\cos\left(\frac{\phi}{f}\right)\right)
\end{equation}
where the branches are labeled by the integer $n$.  As the axion $\phi$ moves through its natural period multiple times the potential changes nonperiodically.  However, when a membrane appears the vacuum jumps to a new branch of the potential labeled by a different value of $n$.  Should tunneling of this nature be very likely in the inflationary regime then the inflaton would quickly tunnel through the branches to the bottom of its potential instead of slowly rolling down.  The end result would be a universe that has inflated for many fewer e-folds than initially expected.  To be safe, the tunneling rate must be highly suppressed.

In the recent literature~\cite{Kaloper:2011jz,Brown:2015iha,Franco:2014hsa,Blumenhagen:2015kja,Ibanez:2015fcv,Hebecker:2015zss}  this tunneling rate has been estimated using results derived by Coleman (and de Luccia in the case of a gravitational theory)~\cite{Coleman:1977py,Coleman:1980aw}.  However, there are reasons to doubt the validity of these results when applied to axion monodromy.  First, tunneling between branches in axion monodromy is not a vacuum decay process; the system tunnels between non-metastable states.  Thus, one must account for the difference in kinetic energy during the tunneling process.  Secondly, the spherically symmetric initial conditions studied by Coleman cannot be extended to a typical inflationary state, which breaks this symmetry in the time direction.  Finally, it is by no means clear that all aspects of the `thin wall' approximation used by Coleman will be valid for axion monodromy.  In particular, we will consider bubbles that may be arbitrarily small.

The main purpose of this work is to derive a new estimate for the tunneling rate in axion monodromy using Euclidean methods.  While we believe that our result is an improvement upon a blind application of the formulas given in \cite{Coleman:1977py}, it is important to note that Euclidean techniques, such as the Hartle-Hawking formalism \cite{PhysRevD.28.2960}, are not on firm mathematical footing and rely upon analytic continuations which are difficult to rigorously establish (see, however, \cite{Barvinsky:1993ne}).  The problem is somewhat more acute when we are discussing tunneling in a rolling vacuum where a derivation of the Euclidean formalism via WKB is lacking.  We will nevertheless {\it{assume}} that the Euclidean formalism is valid, which is supported a posteriori by the reasonableness of the results obtained.  With this caveat, we will compute the dominant exponential suppression factor for tunneling amplitudes in a manner analogous to \cite{Coleman:1977py}.

We emphasize that the issue here is not merely the feasibility of axion monodromy as a model of inflation.  Rather, an underlying question is whether or not  super-Planckian field ranges of any sort are realizable in UV completable effective field theories.  We will work in the framework of effective field theory, treating the parameters $\mu$, $e$, and the domain wall tension $T_{2}$ as inputs from the UV.  When evaluating our general formulas to obtain specific decay rates we will assume the WGC relation $T_{2} \lesssim e M_{p}$ as a benchmark and use the tunneling constraint to restrict cosmological applications of monodromy.  We will find, however, that our results generally give a {\it{less}} stringent constraint on tunneling than the previous estimates.  We thus find that tunneling is not an a priori obstacle for these models when $T_{2} = \mathcal{O}(e M_{p})$.

This paper is organized as follows.  In section 2 we review axion monodromy focusing on the details relevant for our purposes.  In section 3 we examine Coleman's formula and methods and derive of a new estimate for the tunneling rate in axion monodromy.  Afterwards,  section 4 evaluates the tunneling rate in relevant limiting cases.  Section 5 contains a description of our new tunneling formulas applied in the contexts of inflation and relaxation.  In section 6 we briefly discuss the problem of including the effects of gravity on our calculation.  Finally, we conclude in section~7.

\section{Review of Axion Monodromy}

In this section we will review the low energy description of axion monodromy as presented in~\cite{Kaloper:2008fb}, highlighting the features that will be important for the discussion at hand.  The model generates an effective mass for an axion $\phi$ without breaking its shift symmetry $\phi \rightarrow \phi + 2\pi f$ by coupling it to a 4-form flux $F_4=dC_3$.  The action is written as follows
\begin{align}
	\label{LorentzianAction}
	S &= \int d^4x \sqrt{g}\left(\frac{M_p^2}{2}\mathcal{R} - \frac{1}{2}\left(\partial\phi\right)^2 - \frac{1}{2}\left|F_4\right|^2 + \frac{\mu\phi}{4!}\frac{\epsilon^{\mu\nu\rho\sigma}}{\sqrt{g}}F_{\mu\nu\rho\sigma}\right) \nonumber \\
	&+ \frac{1}{6}\int d^4x \sqrt{g} \; \nabla_{\mu}\left(F^{\mu\nu\rho\sigma}C_{\nu\rho\sigma} - \mu\phi\frac{\epsilon^{\mu\nu\rho\sigma}}{\sqrt{g}}C_{\nu\rho\sigma}\right) +M_{P}^{2} \int_{\partial\mathcal{M}} d^{3}x\left(\mathcal{K} - \mathcal{K}_{0}\right) \\
	&+ \frac{ n e}{6}\int d^4x \sqrt{g}C_{\mu\nu\rho}J^{\mu\nu\rho} +T_{2}^{(n)} \int_{\Sigma_3} dV_{3}\nonumber
\end{align}
The last term in the first line is the coupling between the axion and 4-form.  This term will give the axion an effective mass analogous to a $B\wedge F$ term in the St\"{u}ckelberg mechanism.  The second line contains boundary terms for $C_{3}$ and the metric\footnote{This is just the standard Gibbons-Hawking boundary term.} which are necessary for the variational principle to be well defined.  The terms in the third line represent the contribution from a bound state of $n$ membranes with total charge $n e$ and tension $T_{2}^{(n)}$.   In this effective theory these membranes are $3$-dimensional surfaces which occupy a region $\Sigma_3$ in spacetime.

\begin{figure}[h]
	\centering
	\begin{subfigure}{0.48\textwidth}
		\includegraphics[width = \textwidth]{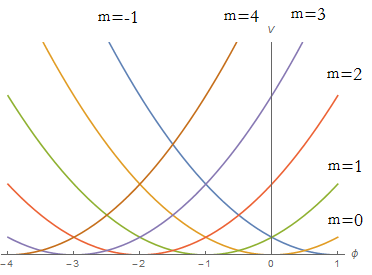}
		\caption{The standard multibranch potential as in~\eqref{EffPot}.}
		\label{fig:EffPot}
	\end{subfigure}
	\begin{subfigure}{0.48\textwidth}
		\includegraphics[width = \textwidth]{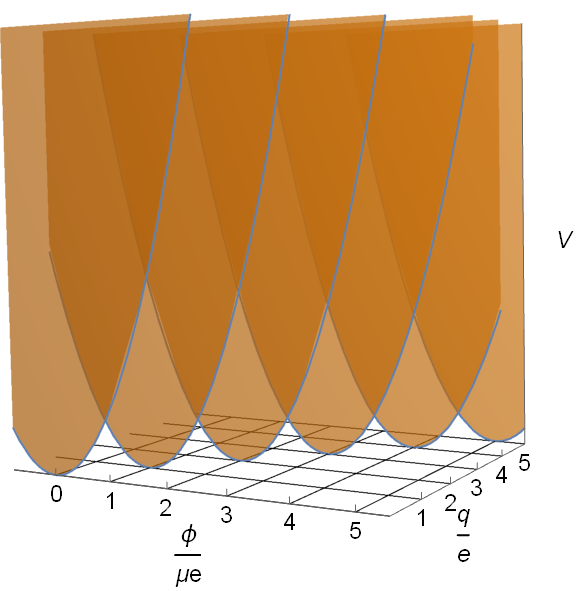}
		\caption{The multibranch potential as described in~\eqref{NewEffPot}.  The shaded regions are above the potential and the white spaces are below the potential.}
		\label{fig:NewEffPot}
	\end{subfigure}
	\caption{Two interpretations of the effective potential.}
\end{figure}

To make the effective mass of the axion explicit we will integrate out $F_4$ in a first order formalism~\cite{Kaloper:2008fb}.  First introduce a Lagrange multiplier to the action
\begin{equation}
	S_q = \int d^4x \sqrt{g}\frac{q}{4!}\epsilon^{\mu\nu\rho\sigma}\left(F_{\mu\nu\rho\sigma} - 4\partial_\mu C_{\nu\rho\sigma}\right)
\end{equation}
Now, upon integrating out $F_4$, the action becomes
\begin{eqnarray}
	\label{KSLag}
	S &=& \int d^4x \sqrt{g}\left(\frac{M_p^2}{2}\mathcal{R} - \frac{1}{2}\left(\partial\phi\right)^2 - \frac{1}{2}\mu^2\left(\phi + \frac{q}{\mu}\right)^2 + \frac{1}{6}\frac{\epsilon^{\mu\nu\rho\sigma}}{\sqrt{g}}C_{\nu\rho\sigma}\partial_\mu q\right) \nonumber \\
	&+& \frac{ n e}{6}\int d^4x \sqrt{g}C_{\mu\nu\rho}J^{\mu\nu\rho} +T_{2}^{(n)} \int_{\Sigma_3} dV_{3}\nonumber
\end{eqnarray}
The quadratic potential for $\phi$ can now be seen explicitly.\footnote{When these setups are embedded into string theory, the potential is typically flattened by corrections that appear at large values of $\phi$~\cite{McAllister:2014mpa,Marchesano:2014mla}. While these corrections have interesting cosmological implications, we will stick for simplicity to the quadratic potential of the effective theory~\eqref{KSLag}.}  As for the Lagrange multiplier, the two terms involving $C_{3}$  in~\eqref{KSLag} show that $q$ only has dynamics inside of the membrane $\Sigma_3$.   On either side of the membrane $q$ will naturally be constant and the shift across the surface determines  the membrane charge
\begin{equation}
	\Delta q = ne \quad\quad n\in\mathbb{Z}
\end{equation}
Now, treating $q$ as taking a discrete set of quantized values, the standard form of the multibranch potential becomes manifest (Figure \ref{fig:EffPot})
\begin{equation}
	\label{EffPot}
	V\left(\phi\right) = \frac{1}{2}\mu^2\left(\phi + \frac{me}{\mu}\right)^2
\end{equation}
The shift symmetry of $\phi$ is preserved by compensating shifts in $m$
\begin{equation}
	\label{KSEffSym}
	\phi \rightarrow \phi + \frac{e}{\mu} \quad\quad m \rightarrow m-1
\end{equation}
and so the potential can be seen as multiple quadratic branches labeled by the integer $m$.

It has been shown in~\cite{Marchesano:2014mla} that~\eqref{LorentzianAction} arises as the low energy effective theory of a wide class of string theory constructions (see also \cite{Blumenhagen:2014gta,Hebecker:2014eua}).  In string theory, the membrane is interpreted as a naturally occurring object, such as a D-brane, charged under some gauge field which compactifies to $C_3$.  One critical feature of the stringy embeddings studied in \cite{Marchesano:2014mla} is that monodromy is generically associated with a discrete gauge symmetry and thus the membranes are charged under a torsion group, $\mathbf{Z}_k$, rather than $\mathbf{Z}$  (see also \cite{BerasaluceGonzalez:2012zn}).  As a result, $T_{2}^{(k)}$ has no topologically protected charge to set a lower bound on the tension.  Nevertheless, we may still recover the limit of small tension, corresponding to an effectively perturbative process,  from the tunneling equations presented below.

We assume the membrane to be thin in the effective theory description.  However, in the full UV theory, the membrane has some thickness controlled by the cutoff $\Lambda \le M_p$.  In this region the field $q$ will have nontrivial dynamics.  This suggests a reinterpretation of the potential in~\eqref{KSLag} as a function of both fields\footnote{Since the UV theory is gauge invariant, the contribution $V_\text{UV}\left(q\right)$ must obey the symmetry~\eqref{KSEffSym}.}
\begin{equation}
	\label{NewEffPot}
	V\left(\phi,q\right) = \frac{1}{2}\mu^2\left(\phi + \frac{q}{\mu}\right)^2 + V_\text{UV}\left(q\right)
\end{equation}
When $q$ is confined by the potential $V_{UV}(q)$ to a local minimum, $q\approx me$, the low energy physics is described by a simple quadratic potential for $\phi$.  However, Figure \ref{fig:NewEffPot} makes it clear that tunneling is possible if the field $q$ climbs over the potential barrier.  The tension $T_2^{\left(1\right)}$ is defined by integrating the potential across its surface such that $V_{UV}^{max} \sim T^{(1)}_{2}\Lambda$.  For our purposes, however, we will not concern ourselves with the detailed form of the potential since we will only need the tension as a UV output.

In addition to the theory as detailed above, axions may couple to instantons that generate a contribution to the potential of the form
\begin{equation}
	\label{InstPot}
	V_\text{inst}\left(\phi\right) = \Lambda_\text{inst}^4 e^{-S_\text{inst}}\left(1-\cos\left(\frac{\phi}{f}\right)\right)
\end{equation}
For the theory to be consistent, the natural periodicity of the axion $\phi \rightarrow \phi + 2\pi f$ must be matched by $k$ shifts in the flux $F_4$~\cite{Kaloper:2011jz,BerasaluceGonzalez:2012zn} leading to the condition
\begin{equation}
	\label{ConsistencyCondition}
	2\pi f = \frac{ke}{\mu} 
\end{equation}
Physically, this means that $k$ of the membranes described above will separate two regions of space which lie in the same position in the axion's periodic domain.  In fact, since the membranes are charged under $\mathbf{Z}_{k}$ rather than $\mathbf{Z}$, the two vacua on either side of a stack of $k$ membranes will share the same quantum numbers, differing only by the value of $\phi$.  Whether or not a bound state of $k$ membranes exists is purely a dynamical question.  For instance, if the prefactor of the instanton potential were sufficiently large (e.g., $\Lambda_\text{inst}^{4} e^{-S_\text{inst}} > \mu^{2}f^{2}$), then one could imagine a process described in the low energy EFT whereby a system tunnelled from one minimum of $V_\text{inst}$ to the next~\cite{Hebecker:2015zss}.  The barrier between these two minima could be interpreted as a chargeless brane with non-zero tension.  However, for smaller $\Lambda_\text{inst}$ there is no guarantee that any barrier between the two branches exist.

\section{Tunneling Rate}

We want to estimate the tunneling rate between potential branches in axion monodromy.  This tunneling rate was estimated in~\cite{Kaloper:2011jz,Brown:2015iha,Blumenhagen:2015kja,Hebecker:2015zss,Ibanez:2015fcv} using formulas derived in field theory by Coleman~\cite{Coleman:1977py} and in a gravitational theory by Coleman and de Luccia (CdL)~\cite{Coleman:1980aw}.

In this note, we will not explicitly calculate the effects of gravity on the decay process, which are believed to be important in certain cosmological setups (see, e.g.~\cite{Ibanez:2015fcv}). Nevertheless, the result with gravitational effects is believed to reproduce the flat space result in the limit $V/M_{P}^{4} \ll 1$, which is the case that we are primarily interested in.  We focus therefore on the field theoretical computation of~ \cite{Coleman:1977py}, which we now briefly review to serve as a basis for comparison.

\subsection{Meta-stable vacuum decay}

\begin{figure}[t]
	\centering
	\includegraphics[width = 0.5\textwidth]{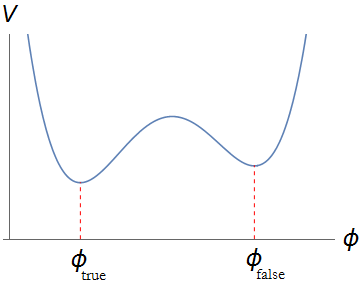}
	\caption{Potential for vacuum decay as studied by Coleman.}
	\label{fig:VacuumDecay}
\end{figure}

The classic reference~\cite{Coleman:1977py} studies the rate of vacuum decay for a scalar field, without gravity, from a false vacuum at $\phi_\text{false}$ to a true vacuum at $\phi_\text{true}$, as in Fig.~\ref{fig:VacuumDecay}.  The Euclidean action is:
\begin{eqnarray}
	S_{E} &=& \int d^4x \left(\frac{1}{2}\left(\partial\phi\right)^2 + V\left(\phi\right)\right)
\end{eqnarray}
and the tunneling rate is estimated as
\begin{equation}
	\Gamma \sim e^{-B}
\end{equation}
where $B$ is the difference of Euclidean action between the bounce solution $\phi_{b}$ and the false vacuum solution $\phi_\text{false}$
\begin{equation}
	\label{BDef}
	B = S_E\left[\phi_{b}\right] - S_E\left[\phi_\text{false}\right]
\end{equation}
The bounce is a solution to the Euclidean equations of motion which mediates the tunneling between the two local minima.  This solution is characterized by a bubble of approximately true vacuum immersed in a sea of approximately false vacuum.  The two regions are separated by a solitonic transition region called the bubble wall.

The bounce must satisfy three conditions~\cite{Coleman:1980aw}:
\begin{enumerate}
	\item{$\phi_{b}\left(x\right)$ must agree with $\phi_\text{false}$ asymptotically (when $r\equiv\left|x\right|\rightarrow\infty$)}
	\item{$\phi_{b} \neq \phi_\text{false}$}
	\item{$\phi_{b}$ must be the solution satisfying the first two conditions that extremizes $B$.\footnote{More precisely, one must find a saddle point of the functional $B$ with precisely one negative mode (see, e.g.~\cite{Weinberg:2012pjx}).}}
\end{enumerate}
In \cite{Coleman:1977py,Coleman:1980aw}, the bounce is calculated under two assumptions: that the bubble size is large relative to the thickness of the wall and that the potential difference between the false and true vacua $\Delta V = V\left(\phi_\text{false}\right) - V\left(\phi_\text{true}\right)$ is small relative to the height of the potential barrier separating the vacua. These rather restrictive conditions are generally referred to as the `thin-wall approximation'.\footnote{The term `thin wall approximation' is not used uniformly throughout the literature (c.f.~\cite{Gen:1999gi,Masoumi:2016pqb}).  The usage here is as defined by Coleman~\cite{Coleman:1977py}.}   The details of the bubble wall are then parameterized by some tension $\sigma$ which contributes to the action.

Both the metric and initial configuration $\phi_i(x)=\phi_\text{false}$ in Euclidean space are $SO\left(4\right)$ symmetric and so the action $B$ is minimized by a spherically symmetric bounce with bubble of radius $\bar{r}$~\cite{Coleman:1977th}.  The action difference~\eqref{BDef} can then be calculated in the thin wall approximation as
\begin{equation}
	\label{ColemanAction1}
	B = -\frac{1}{2}\pi^2\bar{r}^4\Delta V + 2\pi^2\bar{r}^3\sigma
\end{equation}
The first term here corresponds to the energy difference in the interior of the bubble which drives the bubble to expand.  The second term corresponds to the tension of the bubble wall driving the bubble to contract.  Since the stationary points of the action determine the dominant decay channel, $\bar{r}$ must be fixed by extremizing $B$.  This leads to a bubble with radius
\begin{equation}
	\label{ColmanRadius}
	\bar{r} = \frac{3\sigma}{\Delta V}
\end{equation}
and a final action
\begin{equation}
	\label{ColemanAction2}
	B_\text{Coleman} = \frac{27\pi^2\sigma^4}{2\left(\Delta V\right)^3}
\end{equation}

\subsection{Bounces in axion monodromy}
\label{bounces}

The case of interest for us has several key differences relative to the setup reviewed above.  First, tunneling in axion monodromy is not a process of vacuum decay, but rather the tunneling event of interest is between non-metastable states.  As we will see, this difference will have important consequences.  Second, the theories studied in \cite{Coleman:1977py, Coleman:1980aw}  contain only a scalar field, or a scalar field and gravity.  In contrast, axion monodromy also contains the 3-form gauge field $C_3$.\footnote{Vacuum decay in the presence of antisymmetric tensor fields has been studied by Brown and Teitelboim~\cite{Brown:1987dd}, albeit without the coupling to axions crucial for monodromy.}   Finally, the inflationary background does not respect the $SO\left(4\right)$ symmetry assumed above.

On a more pragmatic note, we will not derive the membrane tension from the form of the potential as was done in  \cite{Coleman:1977py,Coleman:1980aw}, rather, the tension,  $T^{(n)}_{2}$, will be treated as an input from some UV completion. In fact, the tunneling direction in field space is not that of the scalar axion at all, but rather the `$q$' direction in Fig.~\ref{fig:NewEffPot}.  As such, the tunneling event studied here more closely resembles the flux tunneling of Brown and Teitelboim~\cite{Brown:1987dd} rather than the vacuum decay of Coleman~\cite{Coleman:1977py}.  However, as the field tunnels in the $q$ direction the dynamics of $\phi$ will be affected due to its coupling to $F_4$.  Since we will be looking at the unadulterated solutions for $\phi$ we are free of the restrictions of the thin wall approximation.  The only assumption we make is that the axion be $\mathcal{C}^{1}$ across the membrane.  To justify this, note that $\nabla^{2} \phi =-V'(\phi)$ and so $\nabla \phi$ is continuous as long as $V'(\phi)$ is bounded.

With these considerations in mind, we proceed to evaluate the tunneling rate in axion monodromy. We follow a procedure similar to that of~\cite{Coleman:1977py} but tailored to the theory of interest. The tunneling rate per unit spacetime volume can be estimated as
\begin{equation}
	\label{gammaB}
	\Gamma \sim e^{-B}
\end{equation}
where $B$ is defined as the difference in Euclidean action between the bounce solution $\phi_{b}$ and some initial solution $\phi_i$ with no membrane
\begin{equation}
	B = S_E\left[\phi_{b}\right] - S_E\left[\phi_i\right]
\end{equation}
Although this Euclidean prescription is difficult to establish rigorously, one may think of Euclideanization as a trick which projects one to the natural `wavefunction of the universe' \cite{PhysRevD.28.2960}.  The factor $e^{-S_{E}(\phi)/2}$ may then be viewed as the amplitude for finding the field configuration specified by $\phi(0)$ and so $\Gamma$ measures the ratio of the probability for tunneling to the probability for no tunneling with the same initial conditions.  This is then interpreted as the tunneling rate.

To construct solutions we begin by Euclideanizing the axion monodromy action~\eqref{LorentzianAction}
\begin{align}
	\label{FullAction}
	S_E &= \int d^4x \sqrt{g}\left(\frac{1}{2}(\partial {\phi})^{2} - \frac{1}{2}\left|F_4\right|^2 + \frac{\mu{\phi}}{4!}\frac{\epsilon^{\mu\nu\rho\sigma}}{\sqrt{g}}F_{\mu\nu\rho\sigma}\right) \nonumber \\
	&+ \frac{1}{6}\int d^4x \sqrt{g} \; \nabla_{\mu}\left(F^{\mu\nu\rho\sigma} C_{\nu\rho\sigma} - \mu{\phi}\frac{\epsilon^{\mu\nu\rho\sigma}}{\sqrt{g}} C_{\nu\rho\sigma}\right) \\
	&+ \frac{n e}{6}\int d^4x \sqrt{g}C_{\mu\nu\rho}J^{\mu\nu\rho} + T_2^{\left(n\right)}\int_{\Sigma_3} dV_{3} \nonumber
\end{align}
From here we will directly integrate out $C_3$ to obtain
\begin{equation}
	\label{C3out}
	S_E = \int d^4x \sqrt{g} \left(\frac{1}{2}\left(\partial{\phi}\right)^2 + \frac{1}{2}\mu^2\left({\phi} + \frac{me}{\mu} - \frac{ne}{\mu}\theta\left(\bar{x} - x\right)\right)^2\right) + T_2^{\left(n\right)}\int_{\Sigma_3} dV_{3}
\end{equation}
where $n \in \mathbb{Z}$ is the charge of the membrane, located at the position $\bar{x}\in\Sigma_3$.  We have already fixed $q$ as appears in~\eqref{NewEffPot} to its integral value as shown in~\eqref{EffPot}.  By integrating out $C_3$ directly the effect of the $n$ stacked membranes on the form of the potential is manifest as a step function.  As expected, the two sides of the membrane exist in different branches of the potential.  Without loss of generality, we will suppress the symmetry manifest in~\eqref{C3out} by setting  $m=0$, which amounts to choosing a particular branch of the potential for the initial state (equivalently, we could reabsorb $m$ into $\phi$ by the redefinition $\phi\to \phi-\frac{me}{\mu}$).
The final action we can then write as
\begin{equation}
	\label{EffAction}
	S_E = \int d^4x \sqrt{g} \left(\frac{1}{2}\left(\partial\phi\right)^2 + \frac{1}{2}\mu^2\left(\phi - \frac{ne}{\mu}\theta\left(\bar{x} - x\right)\right)^2\right) + T_2^{\left(n\right)}\int_{\Sigma_3} dV_{3}
\end{equation}

Now we construct the solutions $\phi_i$ and $\phi_{b}$ from the Euclidean equations of motion.  For our purposes, the initial configuration should be one with no membrane ($n=0$) and the final configuration one with membrane. The equations of motion for each case can be determined easily from~\eqref{EffAction}
\begin{align}
	\label{EOMi}
	\left(\nabla^2 - \mu^2\right)\phi_i &= 0 \\
	\label{EOMf}
	\left(\nabla^2 - \mu^2\right)\phi_{b} &= -ne\mu\theta\left(\bar{x} - x\right)
\end{align}
We construct the final solution from~\eqref{EOMf} according to the conditions for the bounce stated above.  Condition 2 is satisfied automatically as a membrane is included in $\phi_{b}$.  To satisfy condition 1, we will construct the final solution by adding a nonhomogeneous contribution from the step function source to the initial solution that dies asymptotically
\begin{equation}
	\phi_{b} = \phi_i + \phi_s
\end{equation}
This source contribution, $\phi_s$, we can construct with an appropriate Green's function
\begin{equation}
	\phi_s\left(x\right) = ne\mu \int d^4x' \theta\left(\bar{x} - x'\right)G\left(x-x'\right)
\end{equation}
Using this construction of $\phi_{b}$ and the equations of motion, the action difference can be simplified
\begin{equation}
	\label{DiffAction}
	B = \frac{n^2e^2}{2}\int_\mathcal{M} d^4x \sqrt{g} \left(1-\frac{\mu}{ne}\phi_s\right)\theta\left(\bar{x}-x\right) + \int_{\partial\mathcal{M}} d^3x \; n^\mu \sqrt{g} \left(\phi_i\partial_\mu\phi_s\right) + T_2^{\left(n\right)}\int_{\Sigma_3} dV_{3}
\end{equation}
The first term is integrated over the whole Euclidean manifold, $\mathcal{M}$, although only the interior of the bubble contributes.  The second term is integrated on the boundary $\partial\mathcal{M}$ with unit normal vector $n^\mu$.  As we will see this term asymptotes to a constant as $r\rightarrow \infty$.  This represents a considerable departure from the calculation done by Coleman which had no contribution to $B$ from the exterior of the bubble.  The discrepancy becomes clear when looking at the origin of this boundary term: it comes from an integration by parts on the kinetic term of the action.  In contrast, Coleman was considering transitions between static vacua where this term would have been vanishing.

This is as far as the calculation can be taken in generality.  To continue we must study special cases, which we do in the following section.


\section{$SO(4)$ and $SO(3)$ symmetric universes and bubbles}

As can be seen from~\eqref{DiffAction}, one needs only to specify the initial solution $\phi_i$ in order to determine the bounce action.\footnote{ We need to know the membrane shape ($\Sigma_{3}$) as well, but this will ultimately be determined by condition 3 for the bounce.} In addition to choosing an initial solution, there may also be multiple choices of how to analytically continue the Euclidean solution $\phi_{i}$ into a Lorentzian solution. The choice of continuation and the solution together determine the initial conditions in Lorentzian space.  For the sake of reference, we write the Euclidean metric in two different forms which suggest two possible continuations:
\begin{eqnarray}
ds^{2} &=& dr^{2} + r^{2} d\Omega_{3}^{2}   \\ \nonumber
&=&  d\tau^{2} + d\vec{x}^{2}
\end{eqnarray}
One obvious way to perform the continuation is to set $t = -i \tau$  such that $\phi_{i}(0,\vec{x})$ determines the initial conditions on the Lorentzian branch. In order for this patching to make sense we will need to require $\partial_{\tau}\phi(\tau,\vec{x})|_{\tau=0}=0$. We will refer to this as the `flat' continuation.

When using the flat continuation we may consider an ansatz for $\phi_{i}$ preserving either $SO(4)$ (spherical) or $SO(3)$ (cylindrical) symmetry.  For the cylindrically symmetric ansatz it is not immediately obvious what shape the bubble should take so as to extremize the bounce action (\ref{DiffAction}).  However, it is straightforward to check that even when a cylindrical solution is used the boundary term in (\ref{DiffAction}) only depends on the lowest spherical harmonics of $\phi_{i}$ and $\phi_{s}$.  Thus, as a functional of $\phi_{s}$, the bounce action is actually `secretly'  $SO(4)$ symmetric and so it is therefore natural to assume that the bubble should be spherically symmetric as well.  With this in mind, we may proceed to look for spatially homogeneous solutions $\phi_i=\phi_i(\tau)$,  where $\phi_i\left(0\right)$ has a natural interpretation as the inflaton field throughout an infinite spatial slice at time $``t=0"$.  This is the situation that is most relevant physically.  In this case, we will think of the tunneling rate as calculating the probability per unit spacetime of bubble nucleation when the field value is near $\phi_{i}(0)$ throughout space.  This case will be discussed in subsection~\ref{sec:cylindrical}.

If we were to instead consider the $SO(4)$ symmetric solution $\phi_i=\phi_i(r)$ in conjunction with the flat continuation then the initial Minkowski slice would be inhomogeneous.  In fact, this would correspond to highly tuned initial conditions and is not of much physical interest.

A second possible continuation is to think of `$r$' as a time coordinate.  This situation can emerge if one is dealing with a universe with spatial section $S^{3}$.  For instance, if we are continuing from a Euclidean $S^{4}$ with metric $ds^{2} = dr^{2} + \sin^{2}(r)d\Omega_{3}^{2}$ to de Sitter we must analytically continue as $r \rightarrow i t + \pi /2$.  In this case we may choose a Euclidean solution respecting the full $SO\left(4\right)$ symmetry, making it straightforward to determine a unique solution in terms of the parameter $\phi_i(r=0)$. This would give rise to homogeneous initial conditions, though the obvious disadvantage is that this continuation is necessarily tied to universes with $S^3$ spatial sections.

Although we are ultimately interested in the cylindrical solution  continued via $t =- i\tau$, we first investigate  in subsection~\ref{sec:spherical} the computationally simpler case where the full $SO\left(4\right)$ symmetry is preserved by $\phi_{i}$ in order to build intuition.  As mentioned above, this solution may either be interpreted as giving rise to inhomogeneous initial conditions with the flat continuation, or to homogeneous initial conditions on $S^{3}$ using the radial continuation.

\subsection{Spherical Symmetry}
\label{sec:spherical}

We will first consider an initial solution that is spherically symmetric, i.e., $\phi_i = \phi_i\left(r\right)$.  As just discussed, $SO\left(4\right)$ symmetric solutions are only of relevance for closed universes, though it turns out that the result is remarkably similar to the case with cylindrical symmetry studied later.  We begin by writing the spherically symmetric solutions of~\eqref{EOMi} in terms of modified Bessel functions
\begin{equation}
	\label{phiiSphere}
	\phi_i (r) = 2A\frac{I_1\left(\mu r\right)}{r} + \tilde{A}\frac{K_1\left(\mu r\right)}{r}
\end{equation}
For the solution to be nonsingular at the origin we must set $\tilde{A}=0$.  We have included a factor of $2$ so that $\phi_i\left(0\right) = A\mu$. Using the $SO\left(4\right)$ symmetry, the shape of the bubble must be $\Sigma_3 = S^3$ with some radius $\bar{r}$, which is fixed by the requirement that the action be extremized.

In general we can determine $\phi_{b}$ using the Green's function described above, but there is a computationally simpler procedure available because we have closed form solutions to the homogeneous equation~\eqref{EOMi}: find homogeneous solutions in the two regions $r > \bar{r}$ and $r < \bar{r}$ and then match the solutions at the membrane
\begin{align}
	\phi_{b} &=
	\begin{cases}
		\phi_+ \quad r>\bar{r} \\
		\phi_- \quad r<\bar{r} \\
	\end{cases} \\
	\phi_+ &= 2A\frac{I_1\left(\mu r\right)}{r} + c_1\frac{K_1\left(\mu r\right)}{r} \\
	\phi_- &= -\frac{ne}{\mu} + c_2\frac{I_1\left(\mu r\right)}{r}
\end{align}
Just as in~\eqref{phiiSphere}, in the interior region the singular term in $\phi_-$ is dropped.  However, in the exterior region $\phi_+$ it is kept as $\lim_{r\rightarrow\infty} K_1\left(\mu r\right) = 0$.  The coefficient of the first term in $\phi_+$ is then chosen so that $\phi_{b}$ agrees asymptotically with $\phi_i$.  To fix the other two coefficients we must require the solution to be $\mathcal{C}^{1}$ at $r=\bar{r}$, as discussed in Section (\ref{bounces}).   This determines the coefficients as follows:
\begin{align}
	c_1 &= ne\bar{r}^2I_2\left(\mu\bar{r}\right) \\
	c_2 &= 2A + ne\bar{r}^2K_2\left(\mu\bar{r}\right)
\end{align}

\begin{figure}[t]
		\centering
		\includegraphics[width=0.45\columnwidth]{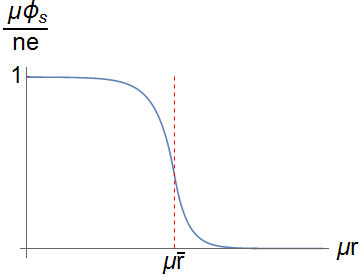}
		\caption{The profile of $\phi_s$ with $\mu\bar{r} = 10$.}
		\label{fig:phis}
\end{figure}

The contribution from the source term is then (see Fig.~\ref{fig:phis})\footnote{Note that the transition region of $\phi_s$ visible in Fig.~\ref{fig:phis} does not represent the bubble wall.  The bubble wall is simply the membrane which then affects the profile of $\phi_b$ via the coupling between $\phi$ and $F_4$.}
\begin{equation}
	\label{phis}
	\phi_s = \phi_{b} - \phi_i = \frac{ne}{\mu}
	\begin{cases}
		\mu\bar{r}^2I_2\left(\mu\bar{r}\right)\frac{K_1\left(\mu r\right)}{r} &\quad r>\bar{r} \\
		1 - \mu\bar{r}^2K_2\left(\mu\bar{r}\right)\frac{I_1\left(\mu r\right)}{r} &\quad r<\bar{r}
	\end{cases}
\end{equation}
It is important to note that $\phi_s$ is independent of the initial configuration and depends only on the shape of the membrane. For any initial solution we choose, an $S^3$ membrane with radius $\bar{r}$ will always give the same contribution to $\phi_{b}$.  This will be a useful result for the cylindrically symmetric initial conditions studied next.

Evaluating the action~\eqref{DiffAction} on our solutions gives
\begin{equation}
	\label{SphereAction}
	B = 2\pi^2\left(\frac{1}{2}n^2e^2\bar{r}^4I_2\left(\mu\bar{r}\right)K_2\left(\mu\bar{r}\right) - Ane\bar{r}^2I_2\left(\mu\bar{r}\right) + T_2^{\left(n\right)}\bar{r}^3\right)
\end{equation}
The final step would be to extremize this action with respect to the membrane radius $\bar{r}$.  Unfortunately, this cannot be done in closed form  so we will have to look at certain limiting cases to estimate the decay rates. We postpone such a discussion to subsection~\ref{sec:limits}.

\subsection{Cylindrical Symmetry}\label{sec:cylindrical}

As mentioned above, the spherically symmetric initial configuration is instructive, but it is not a viable model for axion monodromy in a flat, homogeneous universe.  To interpret our results in the context of standard inflation we should start with an appropriate initial configuration.  We will thus relax to cylindrical symmetry such that $\phi_i = \phi_i\left(\tau\right)$ is only a function of Euclidean time and constant along spatial slices.  Solving~\eqref{EOMi} with this ansatz and the condition $\partial_{\tau}\phi_{i} |_{\tau=0}=0$ gives
\begin{equation}
	\label{phiiCylinder}
	\phi_i\left(\tau\right) = A\mu\cosh\left(\mu\tau\right) 
\end{equation}
Again, the integration constant is defined such that $\phi_i\left(0\right) = A\mu$.

As mentioned previously, one may show by expanding $\phi_{s}$ into spherical harmonics that only the zero angular momentum mode contributes to the boundary term in $B$ while the tension term is manifestly spherically symmetric.  Given this underlying $SO(4)$ symmetry, it is reasonable to assume that the membrane continues to be spherically symmetric.  Fortunately, we already know the source contribution $\phi_s$ in this case~\eqref{phis}.  We then use this to construct the full bounce solution
\begin{equation}
	\phi_{b} = \phi_i + \phi_s
\end{equation}

Evaluating the action using~\eqref{DiffAction} then gives
\begin{equation}
	\label{CylinderAction}
	B = 2\pi^2\left(\frac{1}{2}n^2e^2\bar{r}^4I_2\left(\mu\bar{r}\right)K_2\left(\mu\bar{r}\right) - Ane\bar{r}^2I_2\left(\mu\bar{r}\right) + T_2^{\left(n\right)}\bar{r}^3\right)
\end{equation}
We can see that this takes exactly the same form as the action for the spherically symmetric ansatz~\eqref{SphereAction}.  Thus all of the generic features of this action are the same as in that case.  In retrospect, this result is expected.  In the general form of the action~\eqref{DiffAction} the initial configuration appears only in the boundary term and one can check that only the lowest spherical harmonic contributes.  Thus, there is not much distinction between the spherical and cylindrical solutions.

\subsection{Limiting Cases}\label{sec:limits}

Although it is impossible to solve for the radii $\bar{r}$ that extremize the actions~\eqref{SphereAction} and~\eqref{CylinderAction} in closed form, it is still instructive to look at some limiting cases.  To begin, define the dimensionless parameters
\begin{gather}
	\label{NondimAction}
	B = \left(\frac{\pi ne}{\mu^2}\right)^2\left(\bar{s}^4I_2\left(\bar{s}\right)K_2\left(\bar{s}\right) - \alpha\bar{s}^2I_2\left(\bar{s}\right) + \beta\bar{s}^3\right) \\
	\begin{aligned}
		\alpha &= \frac{2A\mu^2}{ne} \\
		\beta &= \frac{2\mu T_2^{\left(n\right)}}{n^2e^2} \\
		\bar{s} &= \mu\bar{r}
	\end{aligned}
\end{gather}
The parameter $\alpha$ is related to the field configuration during inflation.  We may think of adiabatically scanning through different values of $\alpha$ as we inflate so tunneling needs to be suppressed over a sufficiently large range of $\alpha$.  On the other hand, $\beta$ is fixed by the parameters of the theory and contains information about its UV completion.

We now want to extremize~\eqref{NondimAction} with respect to $\bar{s}$.  The two natural limits to consider are $\beta \ll \alpha$ and $\beta \gg \alpha$, roughly  corresponding to   $\bar{s} \ll 1$ and $\bar{s} \gg 1$.

\subsubsection{Small Bubbles: $\beta \ll \alpha$}

In the limit $\beta \ll \alpha$ the tension is effectively small and we thus expect to have small bubbles ($\bar{s} \ll 1$). In this case, the condition to extremize the action~\eqref{NondimAction} reduces to
\begin{equation}
	\label{SmallCondition}
	3\beta + \frac{1}{2}\left(2-\alpha\right)\bar{s} - \frac{1}{16}\left(4+\alpha\right)\bar{s}^3 = 0
\end{equation}
It is clear that the sign of the term linear in $\overline{s}$ will determine the form of the limiting behavior.  We therefore focus on three separate cases:

\subsubsection*{Subcase I: $\alpha <2$}

In this case, we may neglect the $\beta$ term in equation (\ref{SmallCondition}) and solve.\footnote{In principle, we should further require $\beta \ll (2-\alpha)^{3/2}$}  One finds
\begin{align}
	\bar{s} &= 2\sqrt{\frac{2-\alpha}{3}} \\
	B &= 2\left(\frac{\pi ne}{3\mu^2}\right)^2\left(2-\alpha\right)^3
\end{align}
It is instructive to rewrite these results in terms of the potential difference across the bubble wall along the slice $\tau = 0$, which can be taken as\footnote{This definition is made with specific reference to the cylindrical solutions.  An analogous definition could be made for the spherically symmetric solution, though the interpretation would would have to change accordingly.}
\begin{equation}
	\label{PotDiff}
	\Delta V \equiv V\left(\phi_+\left(\bar{s},\tau=0\right)\right) - V\left(\phi_-\left(\bar{s},\tau=0\right)\right) = n e \mu\,\phi_{b}\left(\bar{s},\tau=0\right) - \frac{1}{2}n^2e^2
\end{equation}
Since we are working with small bubbles, and $\phi_{b}$ is approximately constant near the origin, $\phi_{b}\left(\bar{s}\right) \approx \phi_{b}\left(0\right) = A\mu$ for both initial configurations studied above.  The extremized action becomes
\begin{align}
	\label{NoTensionSmallRadius}
	B &= 2\left(\frac{\pi ne}{3\mu^2}\right)^2\left(1-\frac{\Delta V}{\frac{1}{2}n^2e^2}\right)^{3/2}
\end{align}
Now, in terms of these physical parameters, we can interpret the self consistency conditions for this limit.  Since $\beta \ll 1$, this limit corresponds to $T_2^{\left(n\right)} \rightarrow 0$ and so we do not expect the result to agree with \cite{Coleman:1977py}.  As can be seen from~\eqref{ColemanAction1}, in the limit $\sigma \rightarrow 0$ no bubbles with finite size can nucleate.  In addition, from~\eqref{NoTensionSmallRadius}, we see that $\Delta V < \frac{1}{2}n^2e^2$ in order for the radius to be real.  In terms of the potential in Fig.~\ref{fig:NewEffPot} this limit corresponds to tunneling from one branch to another but ending on the opposite side of the target branch parabola from the starting branch parabola.

\subsubsection*{Subcase II: $\alpha\gg2$}

The behavior of the extremum changes at $\alpha = 2$.  Let's look at $\alpha \gg 2$ for simplicity.  In this case the extremized radius and action become
\begin{align}
	\label{Bag2}
	\bar{s} &= \frac{6\beta}{\alpha} \\
	B &= 27\left(\frac{\pi ne}{\mu^2}\right)^2\frac{2\beta^4}{\alpha^3}
\end{align}
Since we are still working in the small bubble limit we can sensibly rewrite $B$ in terms of the potential difference~\eqref{PotDiff}
\begin{align}
	\label{AMAction}
	B &= 8\frac{27\pi^2\left(T_2^{\left(n\right)}\right)^4}{2\Delta V^3}
\end{align}
This is the limit that most closely resembles the formula derived by Coleman\footnote{Naively, one may think that a comparison with Coleman is not meaningful for small bubbles $\bar{s} \ll 1$ since this violates the thin wall approximation used in~\cite{Coleman:1977py}.  However, the thin wall approximation requires the bubble radius to be large relative to the size of the bubble wall, whereas $\bar{s}$ is defined in terms of the parameter $\mu$.  Thus it is possible for Coleman's formula to be valid in this limit.}
\begin{equation}
	B_\text{Coleman} = \frac{27\pi^2\sigma^4}{2\left(\Delta V\right)^3}
\end{equation}
Strangely, the overall numerical coefficient is different by a factor of 8.  This suggests that the Coleman formula is an underestimate for the action in this limit.  Ultimately, this factor comes from the nonzero kinetic term in the action.

\subsubsection*{Subcase III: $\alpha = 2$}

It is clear that the value $\alpha = 2$ is somewhat special since this is when the second term of~\eqref{SmallCondition} is exactly zero.   In this case the radius and action become
\begin{align}
	\label{B2}
	\bar{s} &= 2\beta^{1/3} \\
	B &= \left(\frac{2\pi ne\beta}{\mu^2}\right)^2
\end{align}
From the definition of $\alpha$ we can rewrite the condition for this case as
\begin{equation}
	\label{fkn}
	\frac{\phi_i\left(0\right)}{2\pi f} = \frac{n}{k}
\end{equation}
Physically, when the axion is $\frac{n}{k}$ fundamental domains away from the minimum of its current branch, then this limit describes the nucleation of $n$ membranes as $\phi$ jumps directly to the minimum of the potential $n$ branches away.  Thus the potential difference becomes $\Delta V = \frac{1}{2}n^2e^2$.  We can then rewrite the action as follows
\begin{align}
	B &= 8\pi^2\frac{\left(T_2^{\left(n\right)}\right)^2}{\mu^2\Delta V} \\ \nonumber
	  &= 16 \pi^{2} \left(\frac{M_{p}}{\mu}\right)^{2}\times \left(\frac{T_{2}^{(n)}}{ e n M_{P}}\right)^{2} 
\end{align}

\begin{figure}[t]
	\centering
	\begin{subfigure}{0.48\textwidth}
		\includegraphics[width = \textwidth]{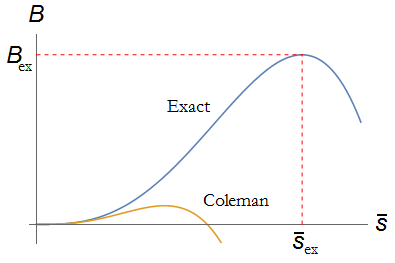}
		\caption{$\bar{s} \ll 1$.}
		\label{fig:SmallBubble}
	\end{subfigure}
	\begin{subfigure}{0.48\textwidth}
		\includegraphics[width = \textwidth]{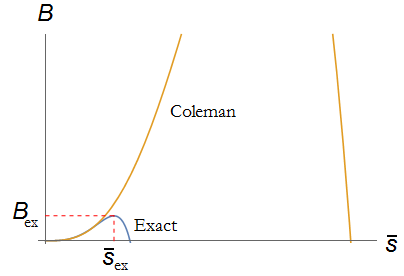}
		\caption{$\bar{s} \gg 1$}
		\label{fig:LargeBubble}
	\end{subfigure}
	\caption{Comparison of the action $B$ between the case studied by Coleman (orange) and the axion monodromy studied here (blue) before extremization with respect to $\bar{s}$.}
\end{figure}

\subsubsection{Large Bubbles:  $\beta \gg \alpha$}

Finally, we consider the case of large bubbles which are found whenever the tension of the domain wall is sufficiently large. Technically, the condition for large bubbles is $\beta \gg \alpha + 1/2$, though we will also assume that $\alpha$ is not too small so that this just reduces to the simpler requirement $\beta \gg \alpha$.  The condition to extremize the action in this limit becomes:
\begin{equation}
	\bar{s}^{1/2}e^{-\bar{s}} \sim \frac{\alpha}{3\sqrt{2\pi}\beta}
\end{equation}
The solution and action can then be estimated as
\begin{align}
	\bar{s} \sim \ln\left(\frac{\beta}{\alpha}\right) \\
	\label{LargeBubbleAction}
	B \sim \left(\frac{\pi ne}{\mu^2}\right)^2\beta\bar{s}^3
\end{align}
In principle one could rewrite these expressions in terms of the jump $\Delta V$ at the membrane, though the result is unilluminating.

\section{Interpretations for the Inflaton and the Relaxion}

We can use these results now in the typical applications of axion monodromy.  In order to derive an estimate in these cases we need a way to constrain the parameters that show up in the action, namely the charge and tension of the membrane.  We will accomplish this by invoking the Weak Gravity Conjecture (WGC)~\cite{ArkaniHamed:2006dz} as was done in~\cite{Brown:2015iha,Ibanez:2015fcv,Hebecker:2015zss}.  The WGC states that for some $p$-form gauge field with $U\left(1\right)$ symmetry group there must exist some object charged under it such that
\begin{equation}
	\frac{T_{p-1}}{e_pM_p} < \left(\frac{T_{p-1}}{e_pM_p}\right)_\text{extremal}
\end{equation}
The right hand side of this condition represents the tension to charge ratio of an extremal $\left(p-1\right)$~brane and is known in generality (see, e.g.,~\cite{Heidenreich:2015nta}).

There are two objects in this theory to which we can apply the WGC: the membrane charged under $C_3$ and the instanton that couples to the axion.  For the latter, the WGC reads
\begin{equation}\label{instantonWGC}
	f\times S_\text{inst} \lesssim M_p
\end{equation}
where the action $S_\text{inst}$ is that which appears in the instanton potential~\eqref{InstPot}.  Although the application of the WGC to instantons and zero-forms is somewhat subtle,~\eqref{instantonWGC} can be related to the more standard WGC for particles and one-forms by arguments involving T-duality and compactification~\cite{Brown:2015iha,Heidenreich:2015nta}.  However, unlike the case of natural inflation, it is not immediately obvious that one must suppress instanton corrections with $S_\text{inst} \gtrsim 1$.  As long as the scale of the instanton $\Lambda_\text{inst}$ is smaller than the scale of the perturbative monodromy potential then $V_\text{inst}$ as in~\eqref{InstPot} will not significantly affect the dynamics of the axion regardless of the size of the action $S_\text{inst}$.  As such, the WGC does not necessarily constrain the decay constant at all in this case.  We will nevertheless assume that the field range $f$ must be sub-Plankian for definiteness.

There is also some debate as to whether the WGC can be applied to a $\left(d-2\right)$~brane such as the one considered in this theory (see~\cite{Heidenreich:2015nta,Ibanez:2015fcv,Hebecker:2015zss}) as there is no corresponding extremality condition for such an object in the absence of a dilaton.  An additional complication for the membranes present in axion monodromy is that they are $\mathbf{Z}_k$ charged.  Since the WGC is defined for objects charged under a $U\left(1\right)$ symmetry, it is not clear that the condition should be applied here.  However, for the purposes of this estimate we will assume the WGC does naively generalize such that
\begin{equation}
	\frac{T^{(n)}_2}{n e M_p} \lesssim 1
\end{equation}
We may also question whether or not the membranes nucleated here are the ones that should satisfy the WGC, but since the decay will be dominated by lighter membranes this is likely to be the case.

Now we may use these constraints in cosmological applications of axion monodromy.  To begin, we rewrite the parameters $\alpha$ and $\beta$ as follows
\begin{align}
	\alpha &= \frac{k}{\pi} \times \left(\frac{\phi_i\left(0\right)}{M_p}\right) \times \left(\frac{M_p}{f}\right) \\
	\beta &= \frac{k}{\pi} \times \left(\frac{T_2^{\left(1\right)}}{eM_p}\right) \times \left(\frac{M_p}{f}\right)
\end{align}
We have rewritten the membrane charge $e$ in terms of the axion decay constant $f$ using the consistency condition~\eqref{ConsistencyCondition}.  We see that the parameter $\alpha$ is proportional to $\phi_i\left(0\right)$, which specifies the field configuration just before tunneling.  In order to test the stability of axion monodromy we must then check the tunneling rate as $\phi_i\left(0\right)$ scans through the slow roll of the axion down the monodromy potential.  The parameter $\beta$ depends on the WGC constraints analyzed above but is itself unconstrained.  It simply depends on the parameters in the theory and can take any value.

\subsubsection*{Axion monodromy inflation}

First we consider the axion monodromy scenario as an inflationary model.  In this case the axion ranges over values from near the beginning of inflation $\phi \gg M_p$ to near the end of inflation $\phi \lesssim M_p$.  We will consider the chance of tunneling in each of these extremes.

Near the beginning of inflation, slow-roll and WGC constraints translate to the condition $\alpha \gg \beta$.  This is the regime in which bubbles are small and the formula~\eqref{AMAction} applies, which agrees parametrically with the formula of Coleman~\eqref{ColemanAction2}.  Near the end of inflation, however, the axion enters a regime where $\alpha \ll \beta$.  This is the condition which leads to large bubbles and the action given by~\eqref{LargeBubbleAction}.  Here the Coleman formula is a bad estimate.  Using this information we can estimate a bound on the allowed field range.  Taking typical values $T_2^{\left(1\right)} \sim eM_p$, $f \sim M_p$, and $\mu \sim 10^{-5}M_p$, the suppression of tunneling $B>1$ requires $\phi \lesssim 10^3M_p$.  This is more than enough to allow the observed $\sim60$ e-folds of inflation.

For $n \ge 1$ the result will depend on $T_{2}^{(n)}$.  If we assume that $T_{2}^{(n)} = n T_{2}^{(1)}$ then it is easy to show that the nucleation process involving a single unit of charge is the dominant decay channel. However, the behavior of the tension as a function of $n$ depends on the detailed UV completion of the theory, and cannot be fully addressed in the effective field theory context of our computations.

\subsubsection*{Axion monodromy relaxation}

We now look at the case of the relaxion~\cite{Graham:2015cka} cast explicitly in the context of axion monodromy~\cite{Ibanez:2015fcv} (see also~\cite{Hebecker:2015zss}).  The relaxion mechanism is used to address the electroweak hierarchy problem by dynamically driving the Higgs mass to a value much smaller than the cutoff for the theory $M$.  This is done in the context of axion monodromy by introducing a coupling between the Higgs and the 4-form flux $F_4$ with coupling $\eta$
\begin{equation}
	\mathcal{L} \supset \eta\left|H\right|^2F_4
\end{equation}
The resulting effective axion monodromy potential becomes~\cite{Ibanez:2015fcv}
\begin{equation}
	V\left(\phi,H\right) = \frac{1}{2}\mu^2\left(\phi + \frac{me}{\mu}\right)^2 + \eta\left(-M^2 + \mu\phi\right)\left|H\right|^2 + \Lambda_\text{inst}^4e^{-S_\text{inst}}\cos\left(\frac{\phi}{f}\right)
\end{equation}
The relaxion $\phi$ begins at some point in its quadratic potential such that the effective Higgs mass squared is positive.  It then rolls down the potential until passing the value $\phi = \frac{M^2}{\mu}$ giving the Higgs a nontrivial vev.  At this point the instanton induced cosine potential becomes large relative to the monodromy potential, $\mu M^2 \sim \frac{\Lambda_\text{inst}^4}{f}$, such that the relaxion quickly comes to rest in one of the instantonic minima.  The end result is a close cancellation such that the effective Higgs mass is much smaller than the cutoff of the theory $M$.

We can now ask about tunneling in this model when the relaxion is rolling in the regime $\phi \gtrsim \frac{M^2}{\mu}$.  In this regime the monodromy part of the potential is dominant and so corrections to the tunneling rate  due to the Higgs coupling should be small.  In order to determine what limit to evaluate the tunneling rate in we should compare the relative sizes of the parameters $\alpha$ and $\beta$.  Using the WGC to interpret the necessary parameters, the authors of~\cite{Ibanez:2015fcv} assume that the membrane is extremal such that $T_2 \sim eM_p$.  Following this estimate, super-Planckian field values for the relaxion give $\alpha \gg \beta$ whereas sub-Planckian values give $\alpha \ll \beta$.  The size of the relaxion in the relevant regime $\phi \gtrsim \frac{M^2}{\mu}$ is ultimately model dependent.  However, in the minimal model studied in~\cite{Graham:2015cka,Ibanez:2015fcv}, in which the relaxion is the standard QCD axion, $\phi$ will always be super-Planckian.  Thus, to estimate the tunneling rate for this model we should use the formula~\eqref{AMAction} which agrees parametrically with Coleman~\eqref{ColemanAction2}.

We can then explicitly constrain the model using the QCD axion as the relaxion with typical values for the QCD axion, $f \sim 10^9\;\text{GeV}$ and $\Lambda_\text{inst} \sim \Lambda_\text{QCD} \sim 200\;\text{MeV}$, along with the conditions $\phi \sim \frac{M^2}{\mu}$ and $\mu M^2 \sim \frac{\Lambda_\text{inst}^4}{f}$ near the end of relaxation.  As done in~\cite{Graham:2015cka,Ibanez:2015fcv}, we test the model by how large we can make the cutoff $M$.  Requiring the tunneling rate to be suppressed $B>1$ then translates into $M \lesssim 10^9\;\text{GeV}$.  One could alternatively require, as in~\cite{Ibanez:2015fcv}, a small total probability of bubble nucleation in the past light cone of a typical observer. The constraint on $M$, however, does not change significantly since this would just demand $B\gtrsim \ln N$, where $N$ is the number of e-folds for which inflation lasts.

The authors of~\cite{Ibanez:2015fcv} note that typical relaxion scenarios lead to bubble radii of order the Hubble radius.  In this case, the effects of gravity become important and our results need to be modified.  As was the case with the results derived in~\cite{Coleman:1977py,Coleman:1980aw}, we expect the gravitationally corrected action to reduce to the flat space solution derived above in the appropriate limit.  Since the flat space action is different from that derived by Coleman, we expect the gravitational correction to also be distinct.  It would be interesting to understand the effects of gravity in more detail, which we briefly discuss next.

\section{Including Gravity}
\label{gravity}

In order to include the effects of gravity we must first find solutions describing the coupled dynamics of $\phi$ and the metric.   The full Euclideanized bulk action after eliminating $C_{3}$ is
\begin{equation}
	\label{GravAction}
	S_E = \int d^4x \sqrt{g}\left(-\frac{M_p^2}{2}\mathcal{R} + \frac{1}{2}\left(\partial\phi\right)^2 + V\left(\phi\right)\right) - M_p^2\int_{\partial\mathcal{M}} \left(\mathcal{K} - \mathcal{K}_0\right)
\end{equation}
with a potential as defined in~\eqref{EffPot}.  The final term of the action is the Hawking-York boundary term which depends on the extrinsic curvature of spacetime, $\mathcal{K}$, and of flat Euclidean space, $\mathcal{K}_0$.

As we saw in the flat space solutions, the type of inflationary initial state we expect produces results very similar to a spherically symmetric initial state.  So we will focus on these types of solutions first for computational simplicity and to gain some insight into the problem.  Take the ansatz
\begin{equation}
	ds^2 = h\left(r\right)^2dr^2 + R\left(r\right)^2d\Omega_3^2
\end{equation}
We choose to work in a gauge where $h\left(r\right) = 1$.  Varying~\eqref{GravAction} with respect to $\phi$ and $h$ gives the equations of motion
\begin{align}
	\label{eomgr}
	\ddot{\phi} + 3H\dot{\phi} &= V'\left(\phi\right) \\
	3M_p^2\left(H^2-\frac{1}{R^2}\right) &= \frac{1}{2}\dot{\phi}^2 - V\left(\phi\right)
\end{align}
where we have defined $H \equiv \frac{\dot{R}}{R}$.  Unfortunately, these equations cannot be solved analytically, even for the relatively simple potential relevant for monodromy.  Instead, one must resort to numerical analysis.

There are several challenges faced when integrating (\ref{eomgr}).  For instance, given typical initial conditions, $R(r)$ is quickly driven to zero both in the forward and backward evolution so that the topology becomes that of $S^{4}$.  However, it is difficult to guarantee that both the geometry and $\phi$ are non-singular at the poles, imposing further constraints on the solutions and requiring an additional regularization procedure (e.g. additional sources).  We hope to explore this further in a future work.

\section{Conclusions}

In this work we have reexamined the process of bubble nucleation in the context of axion monodromy.  Under reasonable assumptions we have derived a result for the decay rate that is parametrically distinct from Coleman's classic result \cite{Coleman:1977py} in several regimes.  However, in the limit most relevant for inflation, i.e.,  $\alpha \gg 2$ and $\beta \ll \alpha$, we find a bounce action which only differs from \cite{Coleman:1977py} by a factor of $8$.  If the brane tension were always near extremal then this result would lead to a highly suppressed decay rate and bubble nucleation would not create an obstacle to inflation.

There are several novel features in the tunneling processes considered here, in contrast to the much studied vacuum decay. Most important is the fact that the scalar field is rolling down a potential and hence has a non-trivial profile (and kinetic energy) even outside of the bubble. As we have seen, this makes a significant difference in the computation of tunneling rates compared to vacuum transitions. We thus expect the type of setup studied here to have wider applicability, especially in the paradigm of eternal inflation in the string theory landscape.

Finally, it is worthwhile to point out that a structure of discrete gauge symmetries relating branches on opposite sides of a domain wall may be a much more general feature of the string landscape.  In the current setup,  the domain walls separating these branches in stringy realizations are $\mathbf{Z}_{k}$ rather than $\mathbf{Z}$ charged. The tension of these (stacks of) branes are particularly important for the proper calculation of the tunneling rate. In the effective field theory approach carried out here, we have used the threshold of extremality and the WGC as a benchmark for estimating the brane tension. There may however be interesting new subtleties associated with such discrete symmetries, especially in concrete string embeddings, that warrant further study.

\subsection*{Acknowledgments}

We would like to thank Daniel Chung, Alex Cole, Inaki Garcia-Etxebarria, Fernando Marchesano, Miguel Montero, Misao Sasaki, and Erick Weinberg for useful discussions. This work is
supported in part by the DOE grant DE-FG-02-95ER40896, the Kellett Award of the University of Wisconsin, and the HKRGC grants HUKST4/CRF/13G, 604231 and 16304414.  G.S. thanks the University of Chinese Academy of Sciences for hospitality during the final stages of this work.

\bibliography{KS_Draft_bib}\bibliographystyle{utphys}

\end{document}